# Investigation of Deformation and Fracture Mechanisms in Two-dimensional Gallium Telluride Multilayers Using Nanoindentation

Yan Zhou[†,*], Shi Zhou[‡], Penghua Ying[#], Qinghua Zhao[§], Yong Xie[∥], Mingming Gong[§], Pisu Jiang[†], Hui Cai[⊥], Bin Chen[⊥], Sefaattin Tongay[⊥], Wanqi Jie[§], Jin Zhang[#,*], Tao Wang[§,*] Dong Liu[†], and Martin Kuball[†,*]

[†] Center for Device Thermography and Reliability (CDTR), H. H. Wills Physics Laboratory, University of Bristol, Tyndall Avenue, Bristol BS8 1TL, UK.

[‡] University of Science and Technology of China, Hefei 230026, China

[#] School of Science, Harbin Institute of Technology, Shenzhen 518055, China.

[§] State Key Laboratory of Solidification Processing, School of Materials Science, Northwestern Polytechnical University, Xi'an, 710072, China.

[∥] School of Advanced Materials and Nanotechnology, Key Laboratory of Wide Band-Gap Semiconductor Materials and Devices, Xidian University, Xi'an, 710071, China

[⊥] School for Engineering of Matter, Transport and Energy, Arizona State University, Tempe, Arizona, AZ85287, USA

[*] Corresponding author. Email address: martin.kuball@bristol.ac.uk, yan.zhou@bristol.ac.uk, jinzhang@hit.edu.cn, taowang@nwpu.edu.cn.

**Abstract:** Two-dimensional (2D) materials possess great potential for flexible devices, ascribing to their outstanding electrical, optical, and mechanical properties. However, their mechanical deformation property and fracture mechanism, which are inescapable in many applications like flexible optoelectronics, are still unclear or not thoroughly investigated due methodology limitations. In light of this, such mechanical properties and mechanisms are explored on example of gallium telluride (GaTe), a promising optoelectronic candidate with an ultrahigh photo-responsibility and a high plasticity within 2D family. Considering the driving force insufficient in atomic force microscopy (AFM)-based nanoindentation method, here the mechanical properties of both substrate-supported and suspended GaTe multilayers were systematically investigated through full-scale Berkovich-tip nanoindentation, micro-Raman spectroscopy, AFM, and scanning electron microscopy. An unusual concurrence of multiple pop-in and load-drop events in loading curve was observed. By further correlating to molecular dynamics calculations, this concurrence was unveiled originating from the interlayer sliding mediated layers-by-layers fracture mechanism within GaTe multilayers. The van der Waals force between GaTe multilayers and substrates was revealed much stronger than that between GaTe interlayers, resulting in the easy sliding and fracture of multilayers within GaTe. This work provides new insights into the deformation and fracture mechanisms of GaTe and other similar 2D multilayers in flexible applications.

## 1. Introduction

Two-dimensional (2D) materials have attracted tremendous interests ascribed to their extraordinary electronic, optical and mechanical properties compared to their bulk counterparts. Recently the preparations of 2D materials through mechanical exfoliation or chemical synthesis have achieved great advances, enabling renewed investigation into 2D materials beyond graphene.[1] Various unique optical and electrical properties have been demonstrated by 2D materials,[2-13] such as high electron and hole mobility (2300 and 1000cm$^2$v$^{-1}$s$^{-1}$ for $\mu_e$ and $\mu_h$, respectively) in multilayered black phosphorus (BP),[14] excellent room temperature current on/off ratio ($10^8$) in monolayer molybdenum disulfide (MoS$_2$) transistors,[15] ultrahigh photo-responsibility ($2\times10^6$A/W) in gallium telluride (GaTe) multilayers.[16-17] In addition, the mechanical properties of 2D materials are also noted crucial for realizing their applications in e.g., flexible, wearable and smart electronics, and have attracted much research interests. Extremely high intrinsic in-plane Young's modulus (~1TPa) and strength (~130GPa) were revealed in graphene using atomic force microscopy (AFM)-based nanoindentation.[18] This AFM-based nanoindentation method combined with molecular dynamics (MD) and other density functional theory (DFT) calculations has been extended to measure the nano-mechanical properties of other 2D materials such as $MoS_2$ and hexagonal boron nitride (hBN).[19-21] Specifically, detailed knowledge about deformation, fracture, generation of defects and potential phase-transition of these 2D materials at nanoscale can be quite different and is essentially required when their devices are under large strain/stress or undergo numerous loading cycles.[20, 22] For example, the strength of graphene can be significantly influenced by out-of-plane deformation under external shear loading[23] or intrinsic topological defects,[24] leading to a failure strength of >50% reduction.

To characterize the local nano-mechanical properties of 2D materials (e.g., in-plane Young's modulus, tensile strength and failure-strain), the aforementioned AFM-based nanoindentation is commonly used.[18, 25-29] However, inaccuracies are noted in this method, mainly due to the fact that the locally concentrated stress near the sharp AFM-tip after crack initiation cannot provide sufficient driving force for further crack propagation.[30] Recently, full-scale nanoindentation techniques with a Berkovich-indenter have been used to measure the mechanical properties and nanometer-scale structural changes in small-volumes of materials and ultrathin-films. In the load-displacement curves produced by this Berkovich-tip nanoindentation method, step-like pop-ins (PI) are often observed which can be ascribed to dislocation-nucleation, slippage, phase-transition, crack-formation, and so on.[31-35] Discontinuous-like push-out (PO) events in the unloading curve are usually reported as indications of phase-transition.[36-37] Load-drop (LD) events (sudden load decrease at a certain displacement) can also be seen, which have been primarily correlated to the formation of interfacial cracks.[34]

When comparing the mechanical properties of reported 2D materials, interlayer interactions are noted playing an important role due to their different strength of van der Waals (vdW) forces existed between individual layers. For example, a 30% decrease in strength is observed in graphene when the number of layers increases from 1- to 8-layers;[26] whereas the strength of hBN is insensitive to layer numbers due to its stronger interlayer vdW-interactions than that of graphene.[26] Interlayer interactions are also found can lead to recoverable sliding within graphene multilayers during the nanoindentation loading.[28] However, the critical mechanical properties and effects of interlayer interactions still have not been investigated in many other 2D materials including the aforementioned GaTe. Specifically, as an exciting emerging material, GaTe multilayers have demonstrated huge advantages with an ultrahigh photo-responsibility for high-performance photodetectors[16-17] as well as giant potential for desirable optoelectronics, electronics and nano-electromechanical system devices with a reported highest anisotropic resistance and a tremendous current on/off ratio within the 2D family;[5, 38-41] these highlighted characteristics make GaTe a great candidate for future nano- and flexible- optoelectronics. GaTe multilayers are also distinguished by an unusually high failure-strain (an AFM-based nanoindentation measured failure-strain of 7%, which is comparable or even better than that of the commonly used PDMS or polyimide flexible substrate).[29] Nevertheless, knowledge about the detailed deformation and fracture behaviors in GaTe multilayers that is crucial for many practical device applications is still lacking.

In this work, the nano-mechanical properties of both substrate-supported and suspended GaTe multilayers are systematically characterized by full-scale Berkovich-tip nanoindentation, micro-Raman spectroscopy, AFM, scanning electron microscopy (SEM) and MD simulations. An unusual concurrence of multiple PIs accompanied with LDs events in loading curves is firstly observed in 2D family, and the mechanisms of interlayer-sliding and layers-by-layers fracture are unveiled and investigated in detail.

## 2. Results and discussion

### 2.1 Nanoindentation characterization of supported GaTe multilayers

The GaTe multilayers samples tested are listed in **Table 1** including the sample thickness, number of layers and the maximum indentation depth. It was found that the load-displacement (*P-h*) curves obtained from different indentation loadings performed on $SiO_2$/Si-substrate-supported samples can be classified into three types; a typical curve from each type is shown in **Figure 1** based on the results from different samples. The nanoindentation schematic and the GaTe atomic structure are shown in insets of **Figure 1**. Type-I curves are common for those at lower displacement (~80nm, sample-1) nanoindentations. This type of curves has only pop-ins (PIs); these curves are relatively smooth with multiple small PI events present where the load remained relatively constant but suddenly increases in displacement (**Figure 1a**). Although very small, these PIs decrease the slope of the loading curves indicating they are associated with plastic-deformation events. Type-II curves are characteristic for larger displacements (~250nm, sample-2) and they have both PIs and push-outs (POs). Some PIs associated with larger steps at higher load (>1200μN), resulted in more pronounced slope changes in the loading curve. Further, a PO appeared in the unloading curve at ~1300μN, **Figure 1b**. Type-III curves are representative of indents collected at higher displacement (~300nm, sample-3) and are even more complex with PIs, load-drops (LDs) and POs all presented. As shown in **Figure 1c**, significant LDs accompanied with large PI events appear during loading, e.g., LDs at (185nm, ~1550μN) and (240nm, ~2100μN), and a PO during unloading again at ~1300μN – similar to the PO observed in Type-II curve in sample-2 (**Figure 1b**). A push-out is often seen as an indicator of a phase-transition while a pop-in is normally correlated with dislocation-nucleation and slip behavior in Si.[37]

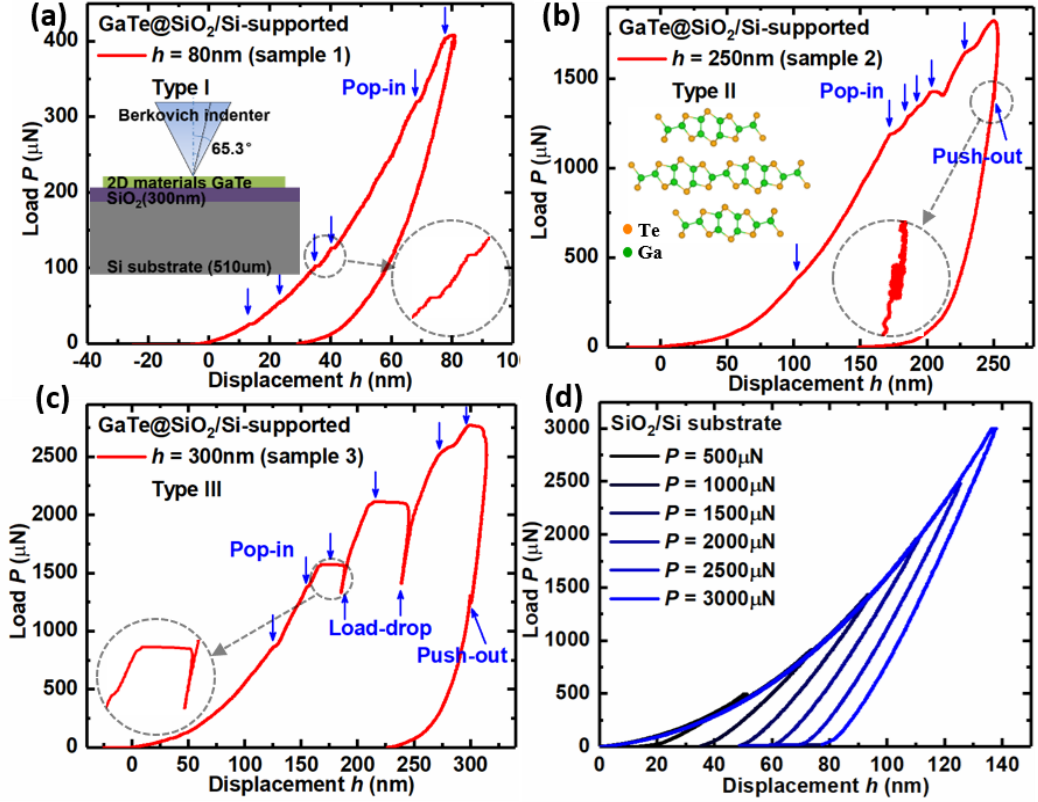

**Figure 1**. Three typical types of load-displacement (*P-h*) curves obtained on SiO$_2$/Si-substrate-supported GaTe multilayers under different indentation depths: (**a**) Type-I (sample-1, *h*=80nm), (**b**) Type-II (sample-2, *h*=250nm), (**c**) Type-III (sample-3, *h*=300nm); (**d**) *P-h* curves of the indents on SiO$_2$/Si-substrate under a series of different loads. Pop-ins (PIs) and load-drops (LDs), push-outs (POs) are labeled in downward- and upward-arrows respectively in all *P-h* curves. Inset in (a) illustrates the schematic of the nanoindentation performed on SiO$_2$/Si-substrate-supported GaTe samples; inset in (b) illustrates the *C2/m*-phase atomic structure of GaTe multilayers; circular insets in (a-c) are enlarged views of the representative PI, PO and LD details.

To examine contributions from the SiO$_2$/Si-substrate itself under the above testing conditions, a series of repetitive indents with increasing peak load were undertaken (**Figure 1d**). It shows that *P-h* curves obtained from the SiO$_2$/Si-substrate are free of PIs, LDs and POs. This confirms that the multiple PIs, LDs and POs in *P-h* curves (**Figure 1a-c**) are genuine reflections of the mechanical responses of the GaTe multilayers. As the nanoindentation depth increases (**Figure 1**), the non-linear loading curve at high-loads/displacements and the increased residual-displacement at complete unloading indicates a change from elastic to inelastic regime accompanied by increasingly pronounced energy dissipating events such as PIs and LDs.

The above indents were compared across different indent-depths; further indents were performed on supported samples at a fixed displacement depth of 200nm in random locations and are detailed in Figure S1-2. All different types of scenarios described in **Figure 1** (small PIs at lower load, large PIs and LDs at higher load and POs during unloading) all repeatedly show up. Despite similar or different film thickness, some PIs (LDs) seem to appear consistently at similar depths/loads. For example, many significant PIs in the 250nm sample (sample-2) are induced at a depth or load that similarly triggers large PIs or LDs in the 300nm sample (sample-3), i.e., about 170nm, 185nm, 210nm, 240nm in depth or 1400μN, 1550μN in load, respectively (**Figure S1**). This implies a sample thickness independent *P-h* curves types and a common material property driven mechanical mechanism exists in GaTe multilayers. Comparing Type-III curves (**Figure S1**), the regularly occurring large LDs indicate multiple fracturing or cracking events in the materials during loading which suggests a transition to irreversible inelastic deformation. Large LDs at lower load, e.g., ~500-μN in sample-5, resulted in a much lower gradient of the loading curve which is a strong implication of modified compliance of the material. We also note that the observed large LDs are spaced at regular loading steps of ~400-700μN, and the accompanied large PI-stage lengths are spaced by ~10-30nm, respectively (**Figure S1-2**). This fact implies multiple step-by-step fracture and sliding behaviors are likely triggered within the GaTe multilayers, which are associated with material properties and require further analysis.

## 2.2 Morphology, microstructure, and micro-Raman stress analysis of nanoindentation

The morphology and microstructure of the indents was investigated by AFM and SEM for further understanding the different types of *P-h* curves. As seen from **Figure 2a-b**, only very small residual-imprints are left after an indentation of 80nm, and the size of the imprints increases with increasing indentation depth. At deeper indentations up to 250nm (**Figure 2c-d**), significant upward deformation of the materials were observed similar to pile-ups geometries caused by plastic-flow in ductile materials. The nature of the upward deformation observed in this work may be different, but we also call it 'pile-ups' in the following. AFM measurements revealed that the heights of the three pile-ups between the sharp corners of the pyramidal indent are ~150nm, 125nm and 100nm, respectively. Three cracks propagating from the corners were observed with a similar length of ~2μm. Notably, one of these cracks was deflected and followed a path parallel or perpendicular to the marked layer-boundaries, indicating that a favored in-plane crack direction probably exists (**Figure 2c**. marked layer-boundary is along the armchair- or zigzag-direction[5, 42]). For the indent produced by an indentation depth of 300nm (**Figure 2e-f**, sample-3), more severe fractures occurred, accompanied with three asymmetric pile-ups whose heights are about 300nm, 50nm and 50nm, respectively. These pile-ups formed between the three corners of the indent. Again there is a favored crack direction perpendicular to the marked layer-boundary direction[5, 42] (**Figure 2e**). Note there is no crack in the direction parallel to the layer-boundary (both planes that parallel and perpendicular to the layer-boundary are preferred in-plane cleavage-planes[42]). These preferred crack propagation orientations indicate the anisotropic nature of the in-plane mechanical properties of GaTe multilayers.

SEM images of other indents performed under the same displacement depth were further analyzed and summarized in **Figure S3**. Similar indent morphologies as that in **Figure 2** are observed. Based on the morphology features and the *P-h* curves (**Figure 1-2**, **Figure S2-3**), the following correlation can be summarized: (i) Pyramidal indent-imprints without macro-cracks, fracture or pile-ups around the indent result in the Type-I *P-h* curve. As there is no fracture or cracks in the residual-imprints (**Figure 2a**), the small PIs at low-load (<500μN) in **Figure 1a** can be correlated to plastic-deformation mechanisms other than fracture and cracks. (ii) Pyramidal indent-imprints with three similar sizes of pile-ups of surrounding materials and macro-cracks lead to Type-II *P-h* curves. More pronounced PIs are caused by the formation of macro-cracks during the indentation loading process, and this resulted in larger permanent residual-displacement upon the reaction of the load. (iii) Pyramidal indent-imprints with fracture, asymmetric pile-ups of the surrounding materials and crack propagations can be correlated to the Type-III *P-h* curve. In this case, the events of larger PIs and significant LDs can be attributed to the formation of obvious fracture in the indent-center. It is worth noting that in the residual-indents, areas with different contrast (**Figure 2c-e**) are consistently observed; these dark areas are likely results of phase-transition of materials driven by the local high-stress generated underneath the indenter.

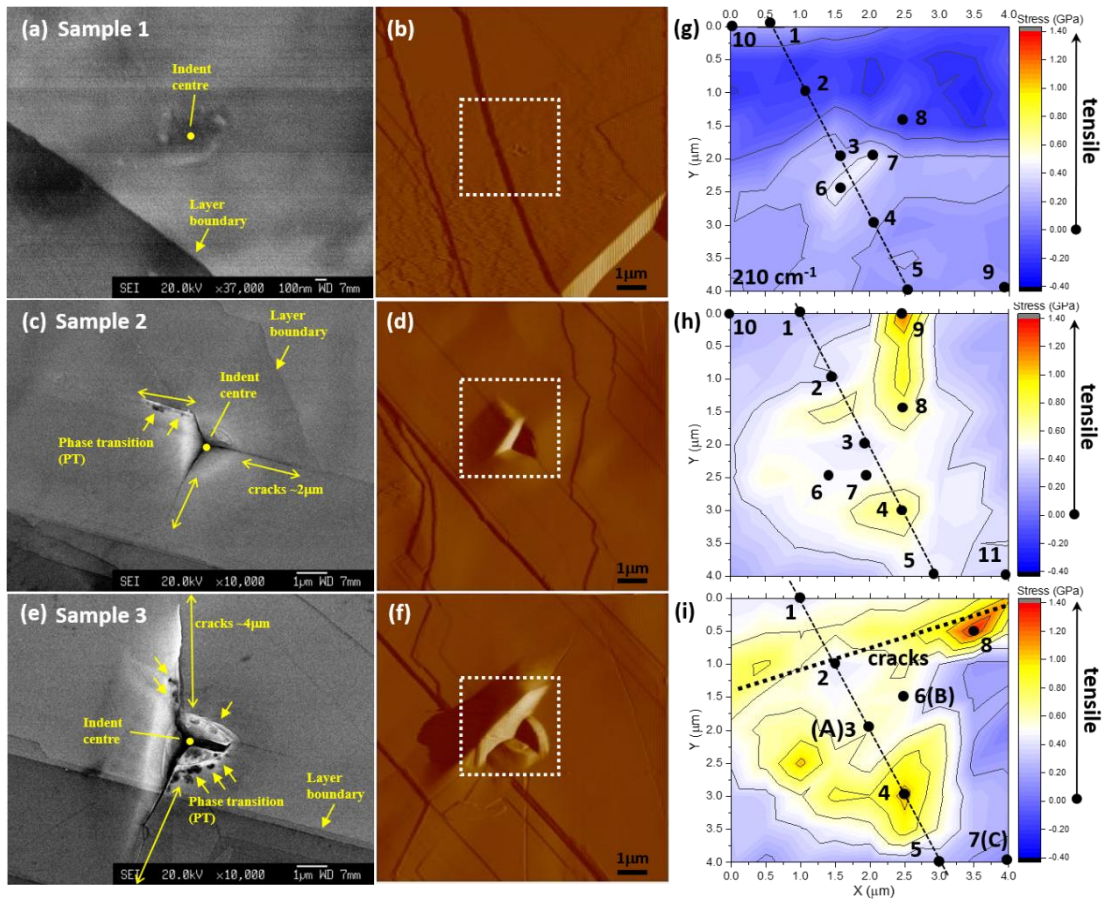

**Figure 2**. SEM and AFM images of the indent morphology for indentation depth of: (**a-b**) 80nm (sample-1); (**c-d**) 250nm (sample-2); (**e-f**) 300nm (sample-3), with the crack lengths, layer-boundaries, indent-center and phase-transition (PT) like dark areas being labeled. The length of the two long cracks in sample-3 reached about 4μm. Stress evaluation mapping of the labeled open dot square area in AFM images after: (**g**) 80nm (sample-1); (**h**) 250nm (sample-2); (**i**) 300nm (sample-3) depth indentation. The mapping area was 4×4μm$^2$, with a step resolution of 0.5μm for all samples. The stress is calculated from the Raman shifts to the reference spectrum based on the stress-sensitive out-of-plane $A_g$ mode (210cm$^{-1}$) using an experimentally obtained stress coefficient of 2.59cm$^{-1}$/GPa; the error bar represents the homogeneity of stress distribution. The dash black line is along one of the axis of the pyramidal indent area and the black dots are selected and numbered for the following Raman spectra comparison (see **Figure S4**).

To further understand the stress distribution after the deformation and fracture process, micro-Raman spectroscopy was used to evaluate and map the residual-stress distribution around the indents. From the indent-imprint created at 80nm, 250nm and 300nm-depth, the residual-stress maps (**Figure 2g-i**) showed an average tensile-stress of about 0.07±0.18GPa, 0.42±0.16GPa and 0.51±0.25GPa, respectively. For the residual-imprint from the 300nm-depth-indent (sample-3, **Figure 2i**), the average residual-stresses formed along the cracks and the edges of the indent-imprints are significantly larger (~0.9GPa) than that away from the cracks (~0.1GPa). The pile-up region exhibits a higher maximum tensile-stress (~1.3GPa) than that of the indent-center (~0.5GPa), but both regions have much higher stress than that of areas away from the indents and cracks (~0.1GPa). Also, there are large variations of residual-stresses over the imprint area, **Figure 2i**. For the residual-imprints of the 80nm- (sample-1) and 250nm-depth-indents (sample-2), a more homogeneous tensile-stress distribution was present around the indent-center, **Figure 2g-h**. The highest tensile-stress in the 250nm-imprint (sample-2) is ~1.0GPa and located in the pile-up region ahead of the macro-crack, similar to the localized stress concentration at the crack-tip in 300nm-imprint (sample-3, **Figure 2i**). For the residual-stress obtained on the same 200nm-depth-indents (sample-4 and sample-5, **Figure S4**), similar average stress of ~0.2GPa and largest tension of ~0.7GPa appear in both Type-II and Type-III indents. To conclude, the average residual-stress around the residual-imprints increased with the indentation depth. In areas away from the indents and cracks, a small tensile-stress of ~0.1GPa exists for all samples. Larger tension is normally found in the pile-up region and the corners of cracks. The indentation process has introduced deformation to the material and as such large localized tensile-stress was formed. This has subsequently resulted in the formation of cracks along the weaker directions. Continued indentation to a higher depth acted as a crack driving force to promote more crack growth and this is evidenced by the higher tensile-stress at the crack tips (**Figure 2h-i**).

## 2.3 Micro-Raman spectrum analysis

Micro-Raman spectra along one axis of triangular imprint from 300nm-indent (sample-3, **Figure 2i**) were inspected in more detail in **Figure 3a**. A significant difference is evident between the indent-center (A), the pile-up region (B) and the area away from the indent and cracks (C). Specifically, broadened Raman peaks associated with amorphization/disorder are observed in the spectrum collected at point-B. Most importantly, several new peaks (e.g., 89, 118cm$^{-1}$) appeared while some originally existing peaks became less apparent (109, 114, 126, 143, 209 and 282cm$^{-1}$) or even disappeared (75, 162, 176, 268cm$^{-1}$) in the imprint region. Similarly, for the 80nm (sample-1), 200 nm (sample-4, 5) and 250nm (sample-2) indent-imprints, new peaks around 90 and 99cm$^{-1}$ are also apparent in the near-indent-center region (**Figure S5-6**); but no significant peak broadening as that of point-B was observed.

Raman spectra with sharp peaks obtained at the indent-center (point-A) after 300nm-depth-indent (sample-3) indicates the amorphization or disorder transition is only localized within the fractured pile-up area; combined with the appearance of multiple LDs, this further implies that the GaTe layers are likely partially fractured layer-by-layer and slide away within the interlayers. We do not exclude the possibility of the formation of new phases during the nanoindentation process induced by the large local stress. Indeed, the areas with darker contrast in the SEM micrographs (**Figure 2c-e**, **Figure 3c**) may originate from changed electrical conductivity in the amorphous-like or new phased transformation materials, similar to those reported in 4H-SiC,[43] GaAs.[44] Notably, the Raman spectra of indent center and darker areas (e.g., point-3, 4, 6) comprise features of both original phase and transformed phase, further indicating these regions are composed of mixed-phases rather than single-phase only.

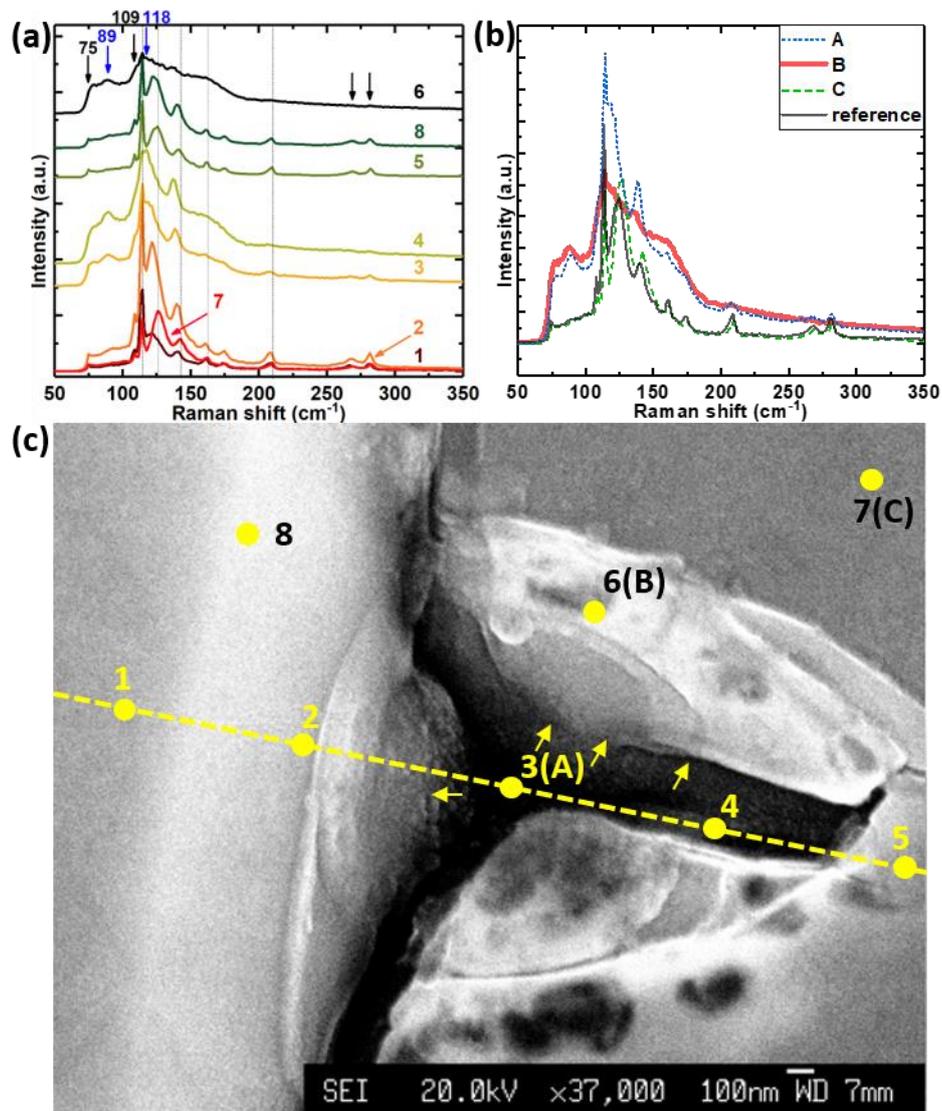

**Figure 3**. GaTe multilayers after 300nm-depth (sample-3) indentation: (**a**) micro-Raman spectra evolution along one axis of the pyramidal indent as the selected points marked in **Figure 2i**; (**b**) detailed Raman spectra comparison between the indent-center and non-indent area, where A and B, C are marked in **Figure 2i**. The reference spectrum was collected before the nanoindentation. (**c**) SEM image of the detailed morphology of indent-center for the sample under 300nm-depth (sample-3) nanoindentation, with the arrows indicating amorphous-like phase-transition areas as evident in the Raman spectra.

**2.4 MD simulation**

To better understand the nanoscale events associated with the indention process, MD simulations were performed for $SiO_2$/Si-substrate-supported GaTe multilayers. Both defect-free and defected GaTe were considered, with the configuration of the model shown in **Figure 4a**. The simulated load-depth curves from three types of thin samples are overlaid in **Figure 4b**. As evident for all three samples, there is a linear-elastic region up to 10nN before LDs occurred and lead the deformation to the inelastic or plastic regime. This is consistent with experimental data. At higher loads, the magnitude of the LDs become more pronounced. Differences to the experiment are due the finite number of layers which can be considered in MD modelling limited by the calculation efficiency and resources, but the main qualitative features from the experiments are reflected in the simulated curves, i.e., concurrence of multiple PIs, LDs and POs. As clearly displayed in **Figure 4c-f** and **Figure S9**, various levels of interlayer-sliding along x-direction accompanied with fracture at different indentation depths are obvious. Therefore, it is apparent from the MD simulations that the original mechanical mechanism of these concurrence events is correlated to layer-by-layer fracture and interlayer-sliding.

Furthermore, it was found that compared with the 10-layer-without-defects sample, the introduction of inevitable void-clusters defects reduced the loading slope after ~5nm-depth and 20nN-load (**Figure 4b**). This is consistent with the 'softening' of the loading curve observed in the experiments (e.g., PI-stages and LDs in **Figure 1b-c**). This also implies the larger PIs and LDs are likely to originate from the multiple layer-to-layer co-fracture and co-sliding within GaTe multilayers facilitated by defects/defects-clusters. For the same indentation depth, the maximum load reached in the 'defected' 10-layer sample (~40nN) has reduced to ~50% of the load in the 'defect-free' 10-layer sample (~80nN). The events of PIs, LDs and POs are all preserved but their magnitude is changed. In the 10-layer-with-defects samples, fracture across multiple layers have been observed and more significant internal interlayer-sliding is present (**Figure 4e-f** and **Figure S9**), corresponding to the relative larger PI-stages in the experiments. Moreover, the simulations also unveil that the vdW-adhesion-force between GaTe multilayers and the substrates (1.83eV/$nm^2$) is much stronger than that between GaTe interlayers (0.56eV/$nm^2$), thus resulting easier fracture of GaTe layers and interlayer-sliding instead of slippage on the substrate.

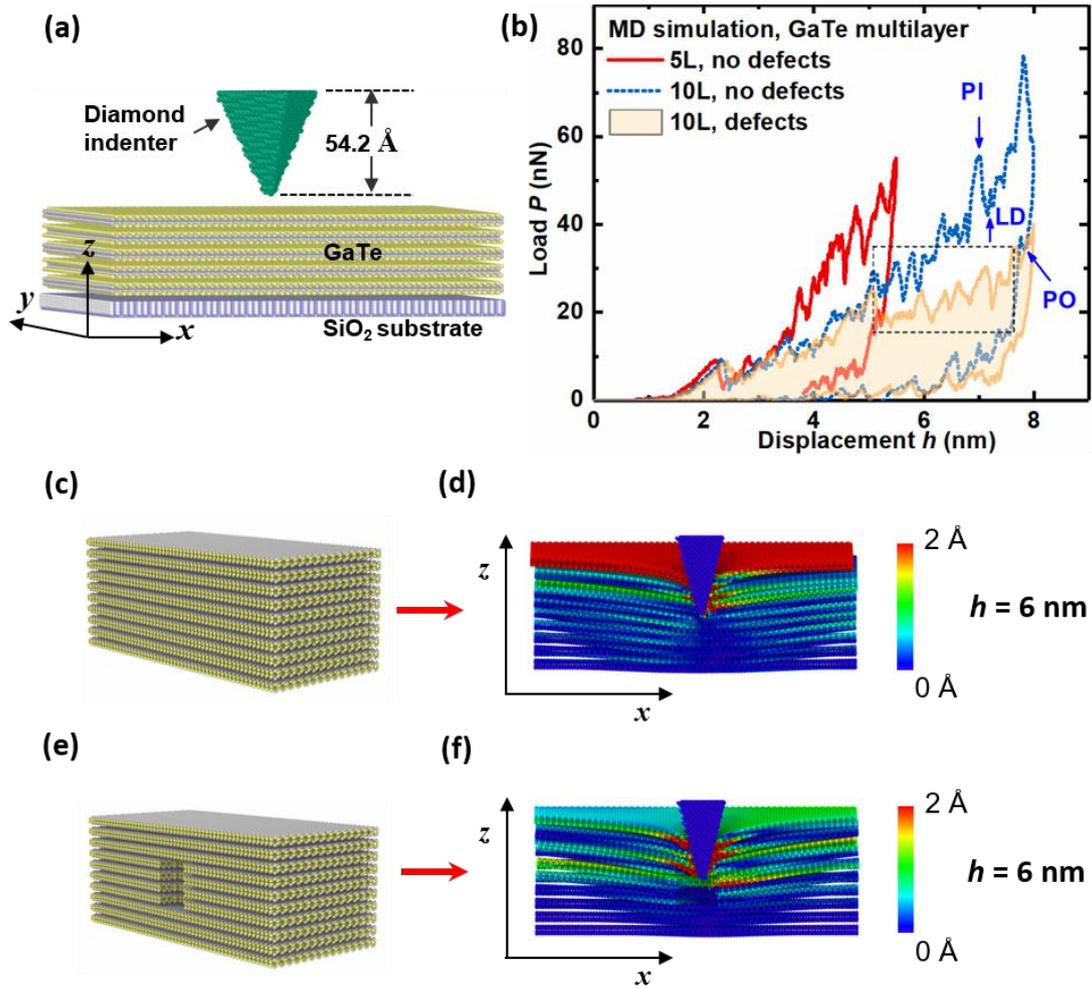

**Figure 4**. (**a**) Sample structure model with a 5-layer GaTe sample of simulated nanoindentation performed by molecular dynamics (MD). (**b**) Comparison of MD simulated nanoindentation curves for 5-layers (5L) and 10-layers (10L) samples with and without the existence of void-defects. Upward and downward arrows represent LDs (PO) and PIs, respectively. (**c**) Sample structure model for '10L, no defects' in MD simulations. (**d**) Fracture and interlayer sliding along x-direction at $h$=6nm for '10L, no defects' obtained in MD simulations. (**e**) Sample structure model for '10L, defects' in MD simulations. (**f**) Fracture and interlayer-sliding along x-direction at $h$=6nm for '10L, defects' obtained in MD simulations. The colourmap represents the different quantity of the atomic displacements during loading process at the indent-depth of 6nm.

## 2.5 Layer-by-layer fracture and interlayer sliding

To provide further evidence for the above mechanical mechanism and exclude any influence from the substrate, nanoindentation experiments were performed on GaTe multilayer samples suspended on rectangular membrane slits fabricated on the SiO$_2$/Si-substrates. Indentation was used to deform the 2D GaTe sample in bending up to 250nm (**Figure 5**). The concurrence of multiple small PIs and LDs, instead of smooth elastic behavior including that at low-load, is observed for all tests (**Figure 5b**). These multiple small PIs events are similar to the behavior reported in single-crystal Pt,[45] i.e., a transition from elastic- to plastic-deformation due to dislocations nucleated under the indenter-tip. Multiple small LDs following some PIs are also evident including at depths of ~120nm, ~180nm and ~240nm, which agree well with those LDs observed in supported GaTe samples (**Figure 1**).

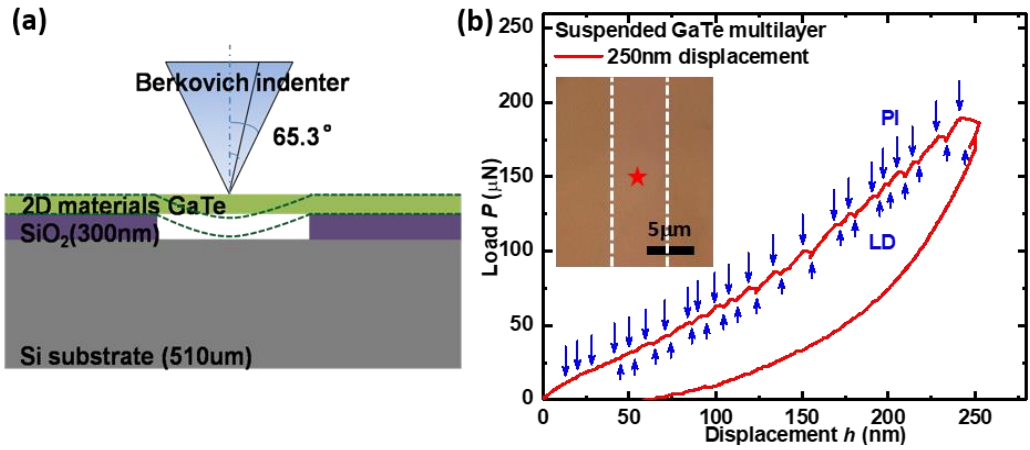

**Figure 5**. (**a**) Schematic of nanoindentation experiment performed on suspended GaTe samples. (**b**) *P-h* curve under 250nm depth (sample 7) indentation for suspended GaTe multilayers. The slits width is ~6μm, and is larger than the cross-sectional tip-radius of ~1087nm at the depth of 250nm (calculated from $R=2h_p\tan 65.3°$, where $h_p$ is the indent-depth and $R$ is the cross-sectional tip-radius at $h_p$). Upward and downward arrows represent LDs and PIs respectively. Inset: Optical microscopy image for one GaTe multilayer sample suspended on a membrane slit in the supporting SiO$_2$/Si-substrate; the mark represents the position where nanoindentation was performed.

However, it is worth noting that no detectable fracture/cracks or a permanent-set indent-imprint could be observed (either by AFM or SEM) even after nanoindentation to 250nm (the maximum available indentation depth in this sample, **Figure S8**). This indicates that the multiple small PIs and LDs are inner features generated within the GaTe multilayers, instead of surface fractures. As the sample was suspended, the maximum load (180μN) was much lower than that of supported samples (1800μN) under the same 250nm displacement. This is plausible as the suspended GaTe films are subjected to mainly bending while the supported films were in more compression as they are supported by a substrate. The absence of large LDs in suspended GaTe films suggests again that those LDs observed in supported films are correlated to the macro-cracks and fractures on the surface.

Performing micro-Raman spectra mapping, similar to **Figure 2-3**, we excluded possible structural changes and stress variations induced during the nanoindentation around the indent-imprints in the suspended films. The measured Raman spectra show no frequency change (**Figure S8**), indicating a less-degraded crystal quality on the sample surface after the nanoindentation than that for the fully supported materials. The Raman spectra evolution along both vertical and horizontal directions of the indent-center shows no other significant difference. This result not only validates the good quality of the crystals in the suspended GaTe after nanoindentation, but also indicates that those PIs and LDs are not attributed to phase-transition or crystal structure changes. As revealed by MD simulations, concurrence of multiple small PIs and LDs in suspended GaTe multilayers are also ascribed to the layer-by-layer fracture, and step-to-step interlayer-sliding behaviors generated within the GaTe multilayers driven by the lateral force from the indenter. Some PIs and LDs are also noted to be a bit larger which can be attributed to defects-mediation.

## 3. Conclusions

The deformation and fracture behaviors of both $SiO_2$/Si-substrate-supported and suspended GaTe multilayers were investigated through full-scale Berkovich-tip nanoindentation, micro-Raman spectroscopy, AFM, SEM and MD simulations. An unusual concurrence of multiple PIs and LDs events is observed in the loading curves of supported GaTe multilayers, accompanied with fracture and cracks. This phenomenon also appears in suspended GaTe multilayers, but without any observable fracture or cracks on the surface. Qualitative MD simulations reveal that such concurrence of multiple PIs and LDs in the loading curve originates from layers-by-layers co-fracture and step-to-step interlayer-sliding within GaTe multilayers during nanoindentation. The vdW-adhesion-force between GaTe multilayers and the substrates is also revealed much stronger than that between GaTe interlayers, thus resulting in the easier fracture and sliding of materials within the GaTe multilayers instead of slippage on the substrate. This work unveils a new deformation and fracture mechanism within multilayered 2D materials, and will underpin device designs, especially for nanoflexible devices based on similar 2D layered materials.

## 4. Experimental details and methods

***Materials and samples preparation.*** The single-crystal bulk GaTe ingot was grown by the modified vertical Bridgeman method, with the help of an accelerated crucible rotation technique to improve the mass and heat transport and smoothen the solid-liquid interface during the crystal growth. High purity powders of gallium (99.99%, Alfa Aesar) and telluride (99.99%, Alfa Aesar) with chemical stoichiometry were mixed in a rocking synthesize furnace and sealed in an evacuated quartz ampoule (<10$^{-4}$ torr vacuums). GaTe flakes were mechanically exfoliated onto 300 nm SiO$_2$/Si or PDMS substrates from a single-crystal bulk GaTe wafer cut from the above GaTe ingot.

*GaTe multilayers transfer and membrane slits fabrication.* To characterize the intrinsic in-plane mechanical properties of both supported and suspended 2D GaTe, free-standing GaTe multilayers were transferred onto SiO$_2$/Si substrates and a series of membrane slits, which were patterned and fabricated onto SiO$_2$/Si substrates via standard CMOS processing, to form supported and suspended samples, respectively (Table S1). After being exfoliated from the bulk using Scotch tape, the GaTe multilayers including tape were attached to a PDMS substrate where thinner GaTe multilayers can be further exfoliated from the tape. GaTe multilayers on PDMS were characterized using micro-Raman spectroscopy at a laser excitation wavelength of 488 nm; no detectable glue residue on the GaTe multilayers was also confirmed using this method. A similar pick-up dry transfer technique[46-48] was developed to transfer the target GaTe multilayers onto designated membrane slits fabricated on the SiO$_2$/Si substrates. The rectangular geometry of these slits was 3-6 μm in width, 20-40 μm in length, and these slits were etched through the whole 300-nm-thick SiO$_2$ films which were thermal oxidized onto the surface of Si substrates.

*Materials properties characterization.* To investigate and understand the mechanical properties of 2D GaTe multilayers, nanoindentation tests were performed using a Hysitron TI 980 Nanoindenter on both supported and suspended GaTe multilayers. A series of indents with different indent depths controlled under the displacement mode were generated using a Berkovich indenter with a 65.3° tip angle. To explore the morphology evolution and visualize the crack details of the indents that were induced by different forces, high-resolution field-emission SEM (Quanta200 FEG, FEI) images were taken. Detailed sample thickness, indentation depths and the tomography of indents were obtained through AFM (Bruker Dimension Edge) using the tapping mode. Residual stress and its distribution after nanoindentation were characterized through the shifts of Raman modes (measured by Renishaw InVia Raman spectrometer at a laser excitation wavelength of 488 nm, using a 100×0.9NA objective with a spot radius of 0.44±0.02 μm), calibrated with the Si Raman line. A reference Raman spectrum of GaTe flakes was measured prior to nanoindentation in the indent area which was later used for the evaluation of residual stress. Raman spectra within the indent region were analyzed for indications of possible phase transformation.

Table 1. List of the GaTe multilayers samples tested and their thicknesses, number of layers and maximum indent depths determined by AFM.

| Sample number | Indent depth, $h$ (nm) | Sample thickness (nm) |
|---|---|---|
| Sample 1 (supported) | 80 | 301.8 (178 layers) |
| Sample 2 (supported) | 250 | 354.3 (209 layers) |
| Sample 3 (supported) | 300 | 372.1 (219 layers) |
| Sample 4 (supported) | 200 | 321.7 (189 layers) |
| Sample 5 (supported) | 200 | 373.8 (220 layers) |
| Sample 6 (supported) | 250 | 779.7 (459 layers) |
| Sample 7 (suspended) | 250 | 779.7 (459 layers) |

*Simulation Methods:* MD simulations conducted in this work were implemented by using the open-source simulation code LAMMPS,[49] in which the standard Newton equations of motion were integrated over time using the velocity Verlet algorithm with a time step of 1 fs. Neglecting the presence of the Si substrate, the simulation box consists of three parts: the diamond indenter, the GaTe multilayers and the $SiO_2$ layer underneath the GaTe (Figure 4 in the main text). Cubic and stishovite crystal structure were adopted for the diamond indenter and the $SiO_2$, respectively. The GaTe structure was obtained from first-principles calculations, in which the bond length $d_{Ga-Te}$ and $d_{Ga-Ga}$ is 2.70 Å and 2.46 Å,[50] respectively. In MD simulations, 5-layered and 10-layered GaTe nanosheets were considered. The spacing of adjacent GaTe layers is 9.2 Å, while the distance between the lowermost GaTe layer and the $SiO_2$ is 5 Å. These values were obtained after a sufficiently long relaxation achieving a convergence in the simulation. The initial distance between the indenter and the uppermost GaTe layer was set as 10 Å to avoid vdW interactions. During the whole simulation process, both diamond indenter and $SiO_2$ layer were regarded as a rigid body by setting the velocity of their atoms as zero. The force interactions between atoms in monolayer GaTe were described by the Stillinger-Weber potential using parameters from Jiang *et al*.[51] The C-C interactions in the indenter were described by the adaptive intermolecular reactive empirical bond order potential,[52] while the interactions between atoms in $SiO_2$ were described by the Tersoff potential[53] using parameters from Munetoh et al.[54] The vdW interactions between adjacent GaTe layers (Ga and Te atoms), the indenter and GaTe multilayers (Ga, Te and C atoms), the GaTe multilayers and substrate (Ga, Te, Si and O atoms) were described by the 12-6 Lennard-Jones (LJ) potential. Detailed parameters used in the LJ potential are listed in Table S2.

Periodic boundary conditions were applied along in-plane $x$ and $y$ directions, while free boundary condition was used along the $z$ direction. Before loading, the GaTe multilayer system was relaxed at 1K in the NVT ensemble (constant atom number, volume and temperature) for 100 ps to obtain an equilibrium structure with a stable energy. After sufficient relaxation, the indenter moves down at a velocity of 0.05 Å/ps during the loading process until achieving the specified displacement or load; then, the indenter moves backwards to the initial position with the same velocity. Note that the force obtained was calculated as the total force on the centroid of the indenter.

**Table S2**. Lennard-Jones (LJ) potential parameters for the nanoindentation MD simulations. Arithmetic mix rule is employed to model the LJ potential between different elements.

| Atom-Atom | C-C [55] | Ga-Ga [56] | Te-Te [57] | Si-Si [58] | O-O [55] |
|---|---|---|---|---|---|
| $\epsilon$ (eV) | 0.00455 | 0.00445 | 0.01727 | 0.01740 | 0.00260 |
| $\sigma$ (Å) | 3.4000 | 1.60000 | 3.98396 | 3.82600 | 3.15000 |

**Supplementary Information**:
The Supplementary Information to this article is available online or from the author.

**Acknowledgements:**

This work was in part supported by the Engineering and Physical Sciences Research Council (EPSRC) under EP/P013562/1. D.L. acknowledges support from EPSRC Fellowship (EP/N004493/2) and New Investigator Award (EP/T000368/1). T.W. acknowledges support by the Shenzhen Virtual University Park under grant of No. 2021Szvup110. Y.X. acknowledges support from the National Natural Science Foundation of China under grant No. 61704129 and No. 6201101295, and the Key Research and Development Program of Shaanxi (Program No.2021KW-02). S.T. acknowledges support from National Science Foundation under grant DMR 1904716, DMR 1552220, DMR 1933214 and the Department of Energy under grant DOE-SC0020653. The authors are grateful to Filip Gucmann (School of Physics, University of Bristol) in helping some measurements of film thickness and morphology using atomic force microscopy. Any opinions, findings, and conclusions or recommendations expressed in this material are those of the authors and do not necessarily reflect the views of EPSRC.

**Conflict of Interest:**

The authors declare no conflict of interest.

**Author contributions:**

M.K. and Y.Z. initiated this collaborative research project. Y.Z. and M.K. designed the experiments. Y.Z. fabricated all the substrates and samples with some help from Y.X., S.Z. and P.J. Y.Z. and D.L. performed the nanoindentation measurements and data analysis. Y.Z. performed the SEM, AFM and Raman measurements and data analysis with help from D.L., S.Z., P.J. and P.T. P.Y. and J.Z. performed the molecular dynamics simulations. Q.Z, T.W., W.J., H.C., B.C. and S.T. synthesized the crystals. Y.Z. prepared the initial manuscript with contributions from all authors to the final manuscript.

# Supporting Information:

# Investigation of Deformation and Fracture Mechanisms in Two-dimensional Gallium Telluride Multilayers Using Nanoindentation

Yan Zhou[†,*], Shi Zhou[‡], Penghua Ying[#], Qinghua Zhao[§], Yong Xie[∥], mingming Gong[§], Pisu Jiang[†], Hui Cai[⊥], Bin Chen[⊥], Sefaattin Tongay[⊥], Wanqi Jie[§], Jin Zhang[#,*], Tao Wang[§,*], Dong Liu[†], and Martin Kuball[†,*]

[†] *Center for Device Thermography and Reliability (CDTR), H. H. Wills Physics Laboratory, University of Bristol, Tyndall Avenue, Bristol BS8 1TL, UK.*

[‡] *University of Science and Technology of China, Hefei 230026, China*

[#] *School of Science, Harbin Institute of Technology, Shenzhen 518055, China.*

[§] *State Key Laboratory of Solidification Processing, School of Materials Science, Northwestern Polytechnical University, Xi'an, 710072, China.*

[∥] *School of Advanced Materials and Nanotechnology, Key Laboratory of Wide Band-Gap Semiconductor Materials and Devices, Xidian University, Xi'an, 710071, China*

[⊥] *School for Engineering of Matter, Transport and Energy, Arizona State University, Tempe, Arizona, AZ85287, USA*

[*]Corresponding author. Email address: martin.kuball@bristol.ac.uk, yan.zhou@bristol.ac.uk, jinzhang@hit.edu.cn, taowang@nwpu.edu.cn.

**Keywords**: Two-dimensional layered materials, gallium telluride, mechanical deformation, fracture, nanoindentation, interlayer sliding.

**Table of contents:**

1. Supplementary nanoindentation characterization and comparison

2. Supplementary morphology and microstructure analysis of nanoindentation

3. Supplementary micro-Raman spectrum and stress analysis of nanoindentation

4. Supplementary AFM and micro-Raman spectrum of suspended samples

5. Supplementary Molecular Dynamics simulations

# 1. Supplementary nanoindentation characterization and comparison

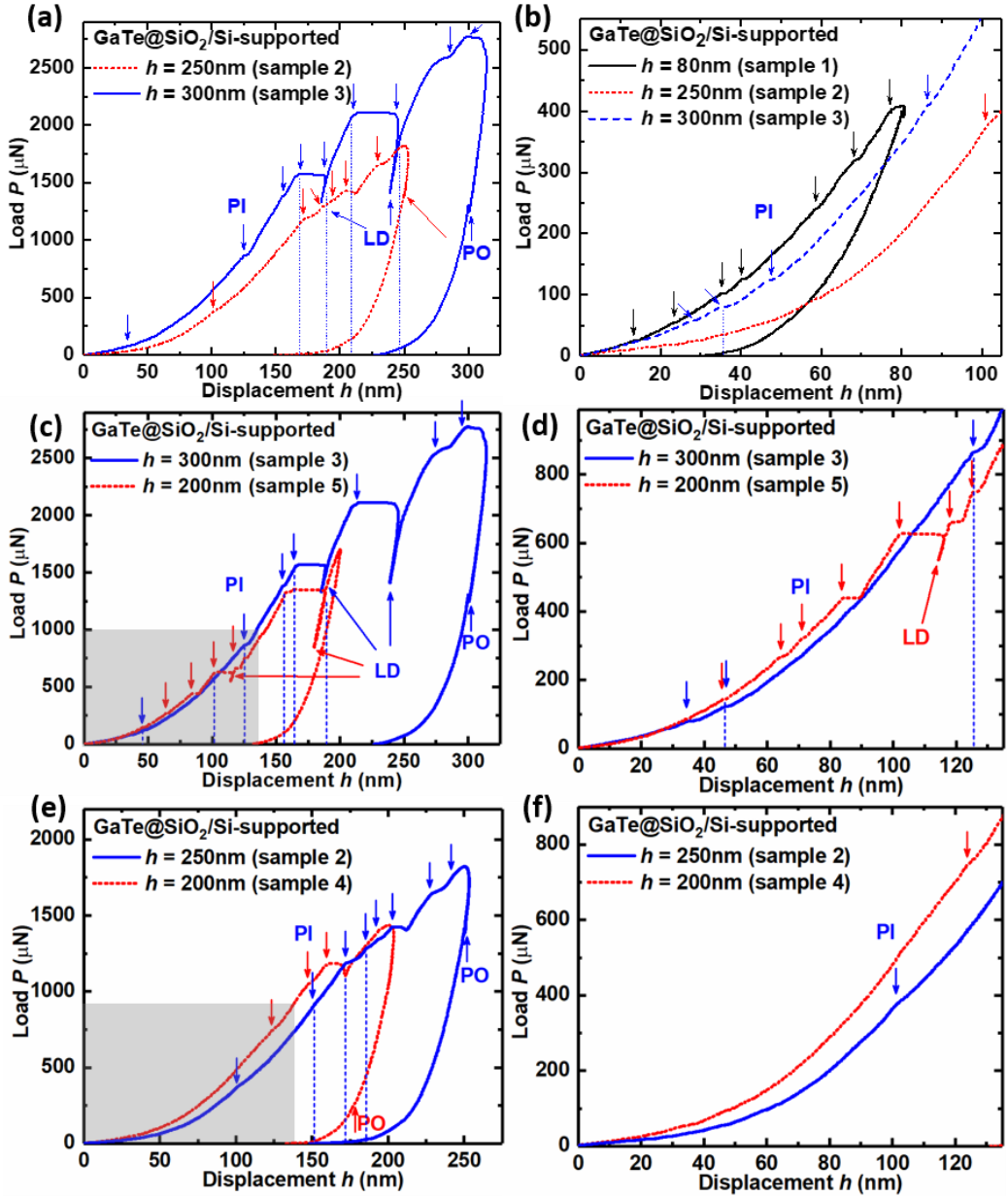

Figure S1. Comparison of *P-h* curves for samples: (a) at indentation depth of *h*=250nm (sample-2) and *h*=300nm (sample-3); (b) details of lower loading region (0-105nm) for *h*=80nm (sample-1), *h*=250nm (sample-2) and *h*=300nm (sample-3) samples. (c) at indentation depth of *h*=300nm (sample-3) v.s. *h*=200nm (sample-5) for Type-III, which with almost the same layer thickness (see Table 1, only one layer difference, ~1.7nm for each monolayer); (d) details of lower loading region (0-135nm) of curves in (c), shaded region; (e) at indentation depth of *h*=250nm (sample-2) v.s. *h*=200nm (sample-4) for Type-II, which with different layer thickness, and (f) details of lower loading region (0-135nm) of curves in (e), shaded region. Pop-ins (PIs) are labeled in downward arrows while load-drops (LDs) and push-outs (POs) are labeled in upward arrows respectively in all *P-h* curves; the vertical dash lines mean PI (LD)

on the compared *P-h* curves appeared at similar indentation depth.

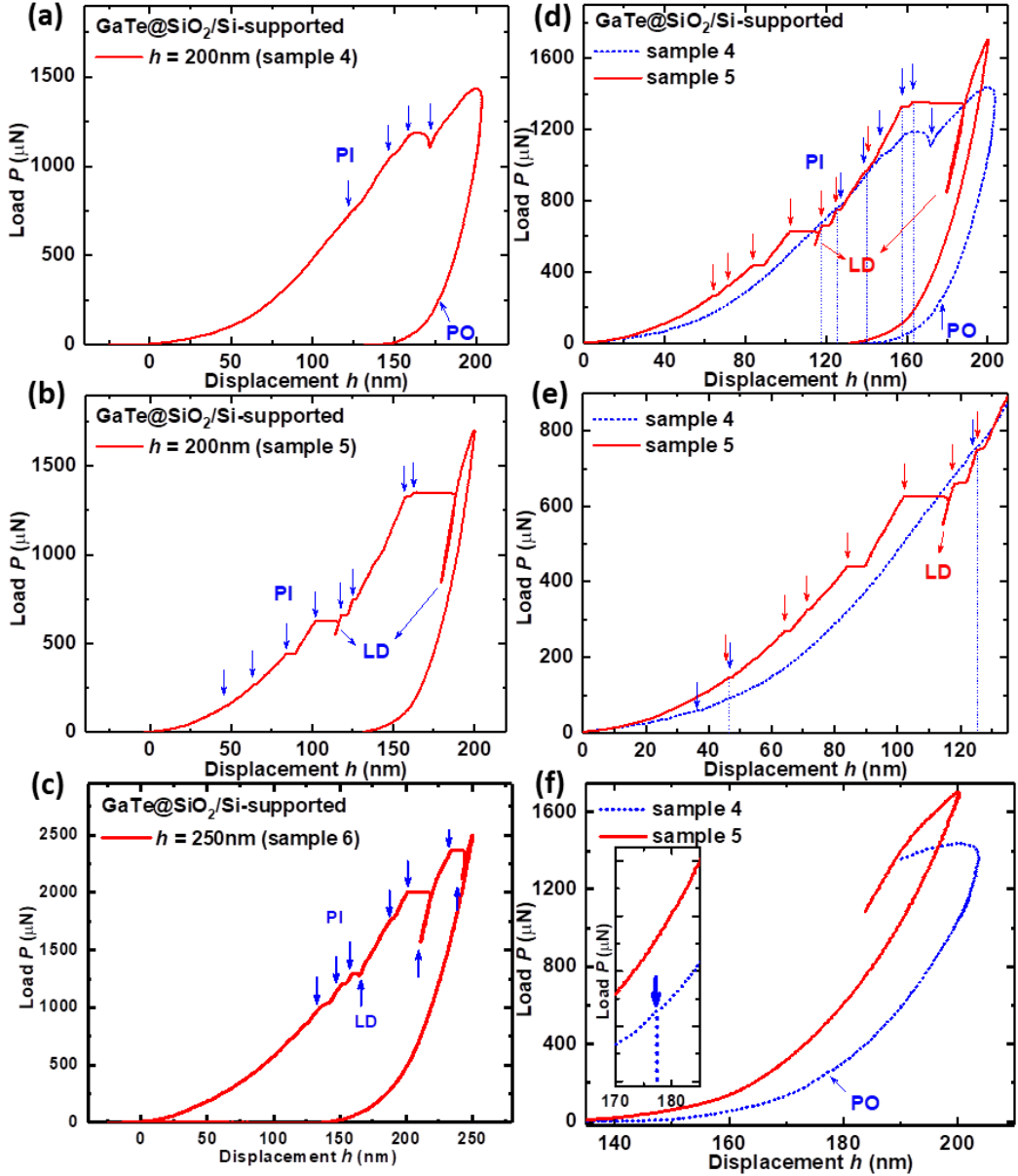

Figure S2. *P-h* curves obtained under the same indentation depth of 200nm in the displacement-controlled mode for: (a) sample-4 (Type-II), and (b) sample-5 (Type-III). (c) *P-h* curves obtained under the same indentation depth of 250nm in the displacement-controlled mode for sample-6 (Type-III). (d) Comparison of *P-h* curves for sample-4 and sample-5, with their lower loading region (<135nm) and unloading region details shown in (e) and (f), respectively. PIs and LDs (POs) are labeled in downward and upward arrows respectively in all *P-h* curves.

## 2. Supplementary morphology and microstructure analysis of nanoindentation

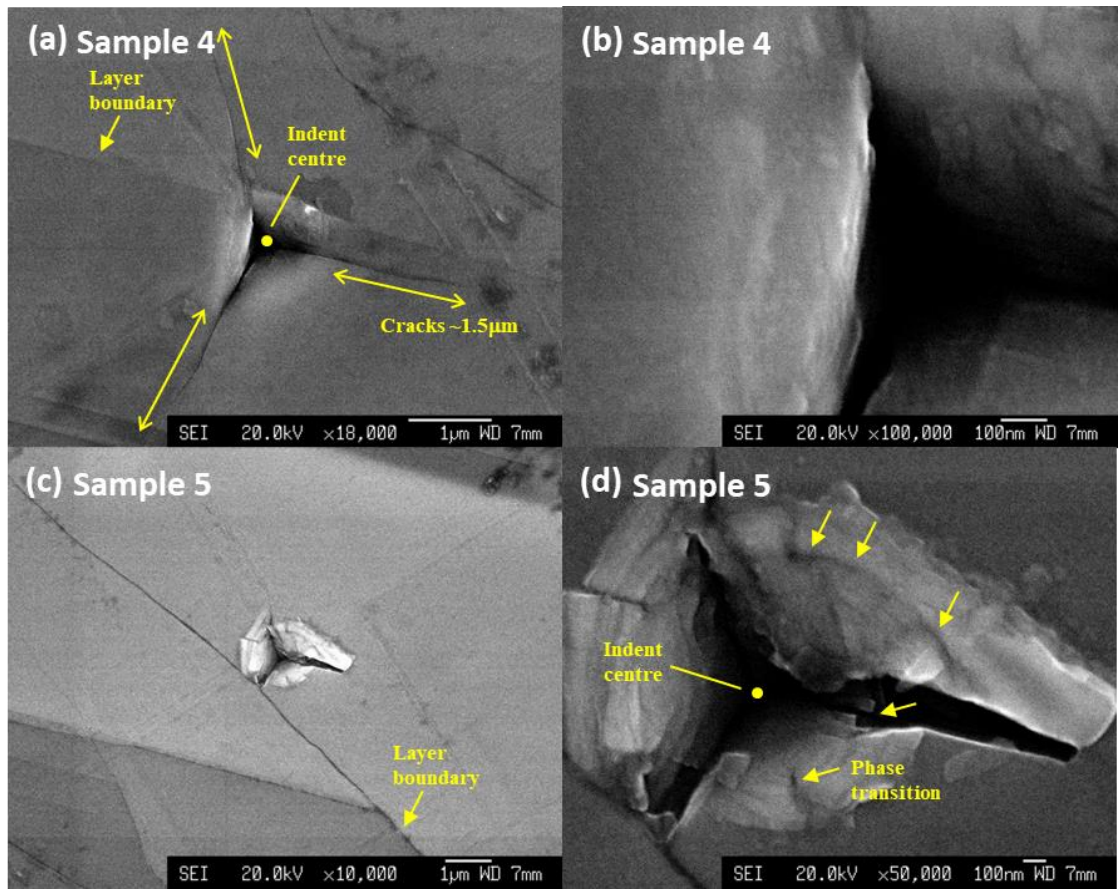

Figure S3. SEM images of the residual indent imprints under a same depth indentation of 200nm, with the cracks-lengths, indent-center, layer-boundaries and phase-transition features being labeled. Image pairs of (a, b) and (c, d) correspond to sample-4 and sample-5 shown in Figure S2(a) and (b), respectively. In sample-4, three almost symmetric pile-ups around the pyramidal imprint were observed accompanied by three similar cracks with lengths of ~1.5μm. In sample-5, the pile-ups are more pronounced but there is no obvious formation of cracks. It should also be noted that only in sample-5, some weak 'darker' fractural features of materials as those in Figure 3-4 were observed.

## 3. Supplementary micro-Raman spectrum and stress analysis of nanoindentation

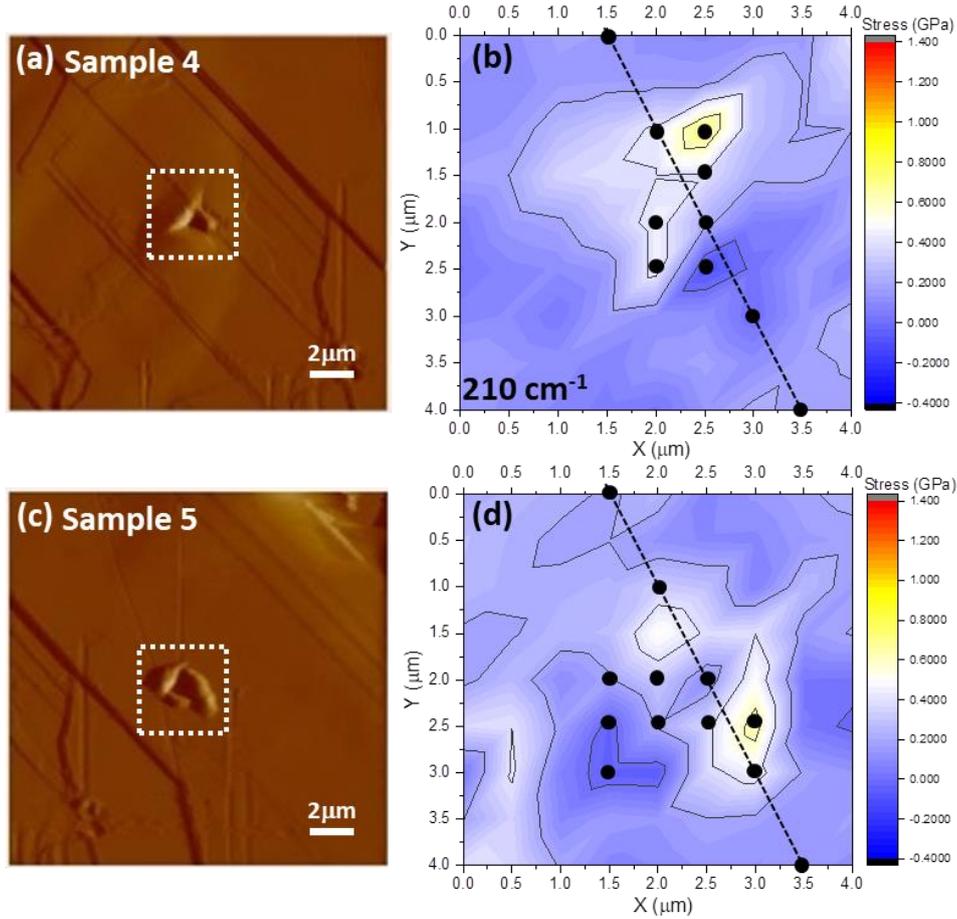

Figure S4. AFM images and Raman mapping area (4×4μm$^2$, indicated by open dot square) of samples after nanoindentation under 200nm-depth displacement load for: (a, c) sample-4 and sample-5, respectively. (b, d) Stress evaluation mapping of the labeled open dot square area for sample-4 and sample-5, respectively. The mapping area has a step resolution of 0.5μm for all samples. The stress is calculated from the Raman shifts to the reference spectrum based on the stress-sensitive out-of-plane $A_g$ mode (210cm$^{-1}$) using an experimentally obtained stress coefficient of 2.59cm$^{-1}$/GPa. The black line is along one of the axis of the pyramidal indent area and the black dots are selected for Raman spectra comparison (see Figure S6).

Figure S4(b, d) shows an average residual stress of about 0.18±0.12GPa (tensile) and 0.20±0.13GPa (tensile) was created in sample-4 and sample-5 after the nanoindentation, respectively (the error bar represents the homogeneity of stress distribution). It is also shown that fracture tends to result in larger tensile residue stress in the indent area, while compressive stress tends to form around the indent to balance the tensile stress generated on the indent. A larger inhomogeneous stress was formed around the edges of the indent imprint in sample-5 (see Figure S4d), similar to the condition of 300 nm depth indent sample (sample-3, see Figure 2i in the main text)

which has the same thickness; this is mainly due to asymmetric or un-sharp crack prolongations formed with the indent fracture thus resulting in an asymmetric stress accumulation.

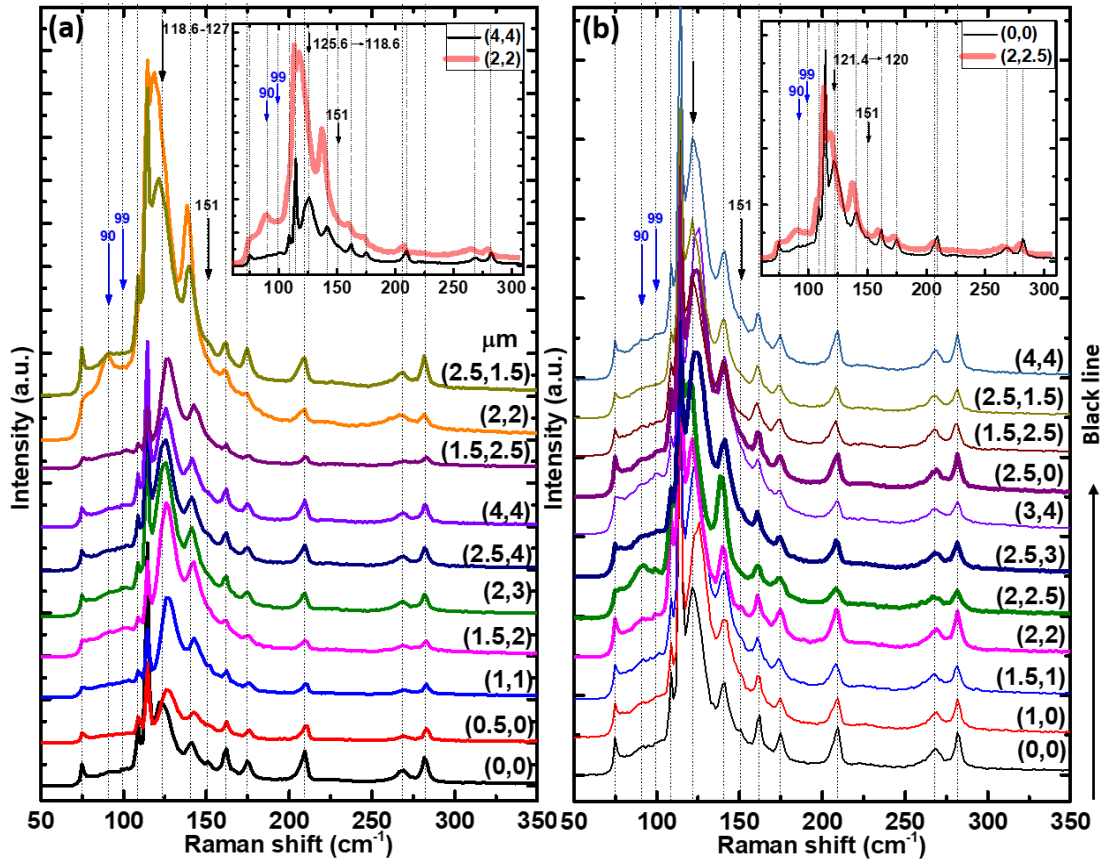

Figure S5. Micro-Raman spectrum evolution along the black line and selected points (as marked in Figure 2 in the main text) after: (a) 80 nm depth indentation (sample-1, corresponding to Figure 2g in the main text), and (b) 250 nm (sample-2, corresponding to Figure 2h in the main text); inset is the detailed Raman spectrum comparison between the indent-center and non-indent area. No significant amorphous-like broadened peaks appeared in the Raman spectrum, while the new peaks around 90 and 99cm$^{-1}$ observed in the near-indent-center region (as shown in insets) which are similar to those features discussed in the main text, likely the consequence from a local amorphization like structure transformations.

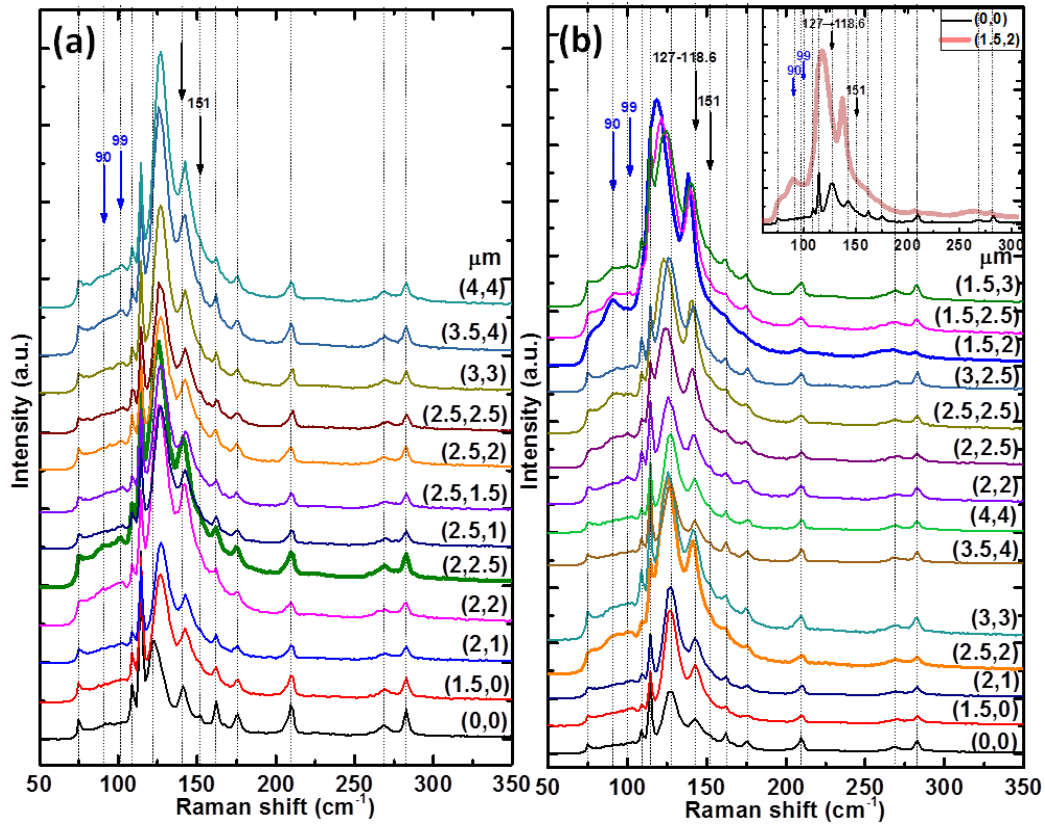

Figure S6. Micro-Raman spectrum evolution along the black line and selected points marked in Figure S4 for (a) sample-4, and (b) sample-5. Inset in (b) is the detailed Raman spectra comparison of the near-center region and the non-indent region; no significant materials changes happened, while an amorphous like Raman spectrum similar to that of Figure 3 and Figure S5b appeared, indicating a local amorphization like structure transformation was similarly induced.

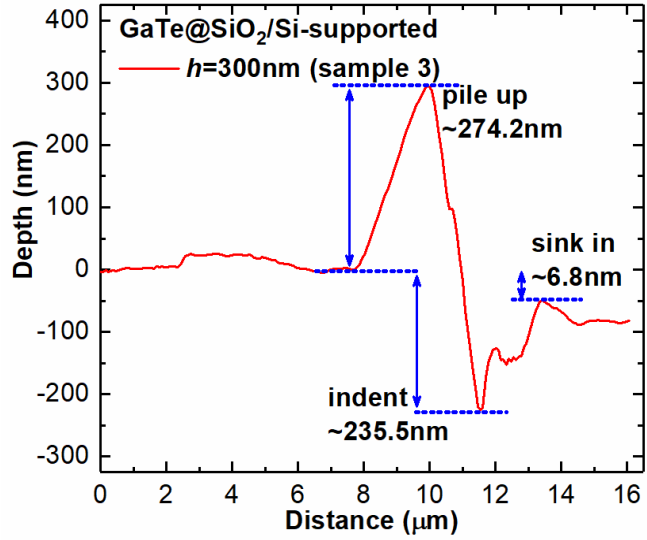

Figure S7. AFM profile measurements on the nanoindentation pit of the 300nm depth (sample-3) indentation sample, with the indent depth, pile up depth and sink in depth labeled, respectively. An indent depth of ~235.5nm was left after the nanoindentation although a 300nm displacement was loaded, and a maximum pile-up of ~274.2nm was resulted while the opposite crack corner presented a slight sink-in depth of ~6.8nm.

## 4. Supplementary AFM and micro-Raman spectrum of suspended samples

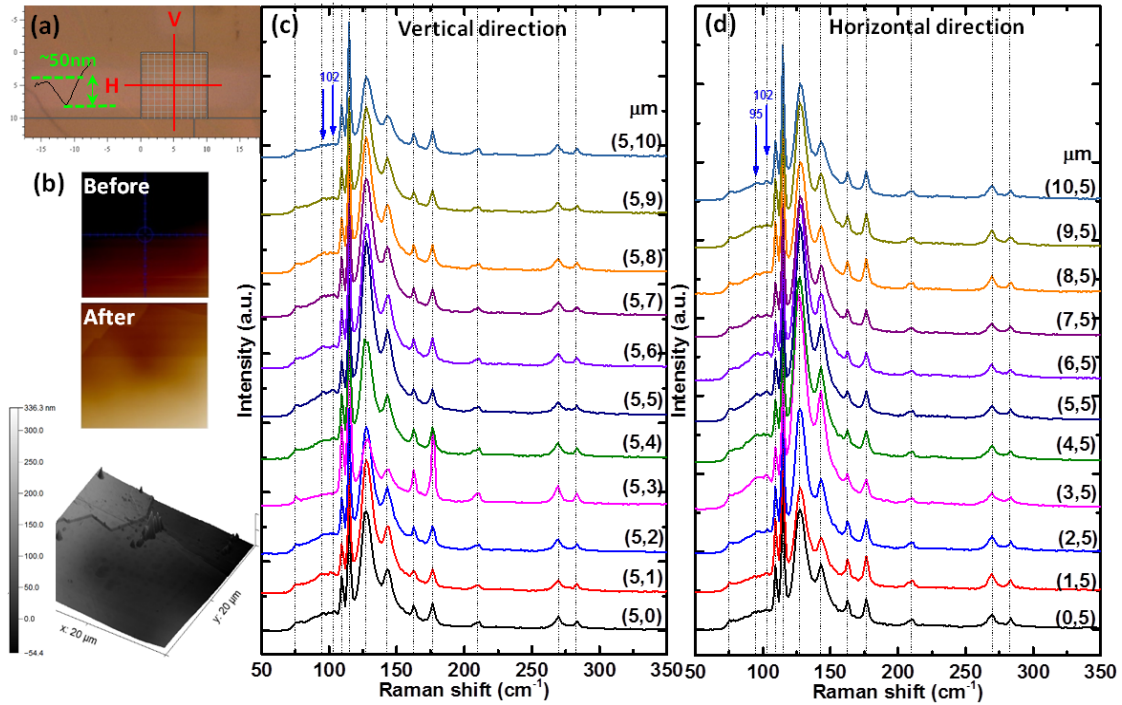

Figure S8. Measurements after 250nm depth (sample-7) indentation on the GaTe multilayers suspended on rectangular membrane slits fabricated on the $SiO_2$/Si substrate: (a) Optical microscopy image and mapping area; (b) AFM images of the indent topography - top figures indent area phase images before and after the indent, while bottom figure is the 3D topography image of indent area; (c) micro-Raman spectra evolution along the vertical direction (V) and (d) horizontal direction (H) of the mapping area.

From the AFM topography, by plotting the depth profile across the indent region, a permanent concave imprint of ~50nm in depth was left after the nanoindentation, as illustrated by the AFM curve in the inset of Figure S8a and the AFM phase images in Figure S8b, indicating a permanent plastic or unrecoverable deformation. Notably, the asymmetric depth profile may be due to the indentation position not at the center of the rectangular slits (located at ~4.5μm/6μm position of the slits). No observable difference can be seen in the Raman spectra, implying an unchanged sample quality even after nanoindentation of an available maximum depth (250nm in this work).

## 5. Supplementary Molecular Dynamics simulations

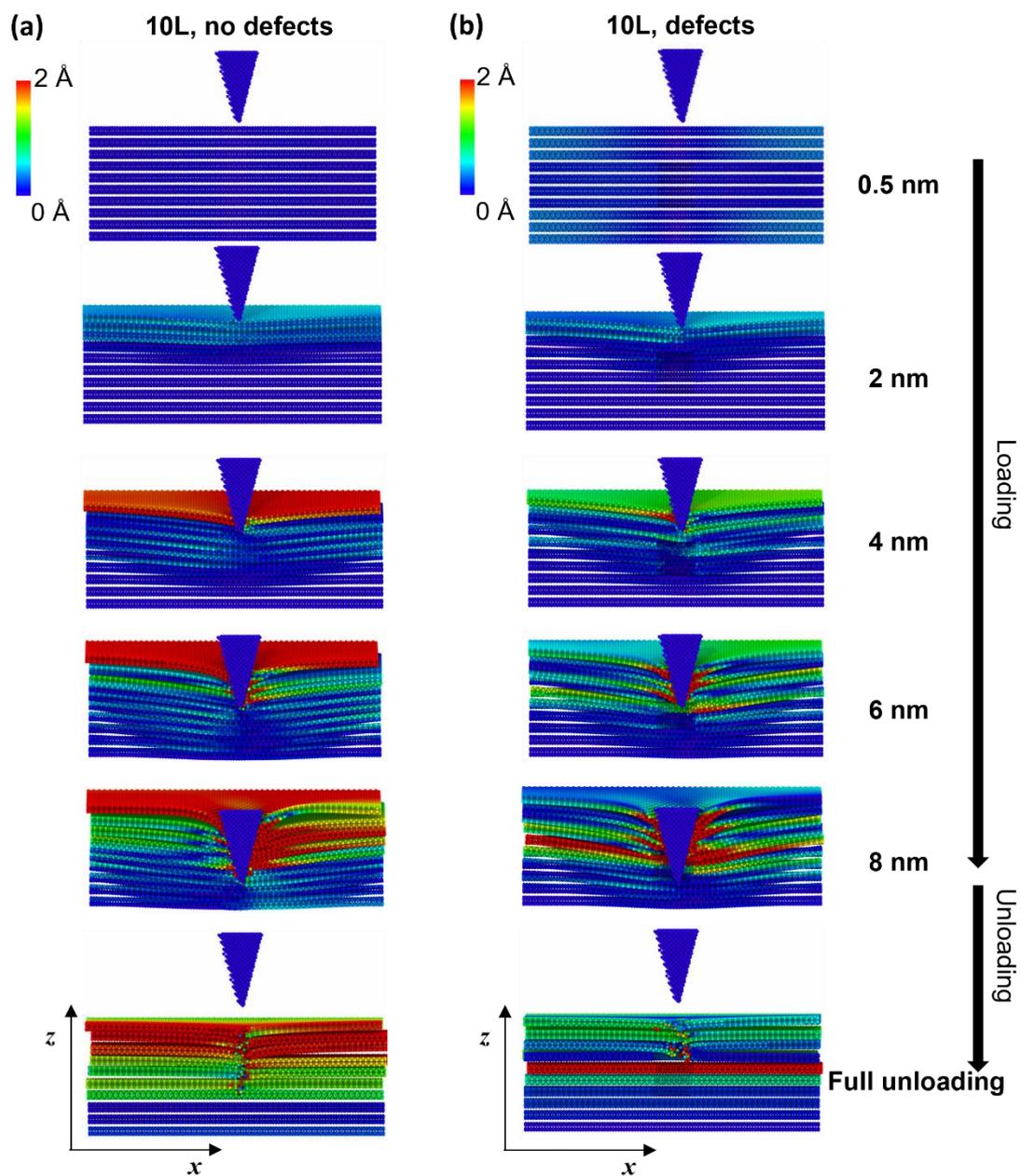

Figure S9. Details of the generated fractures and interlayer sliding along the in-plane $x$-direction at various depth of loading and unloading process for (a) '10L, no defects' samples and (b) '10L, defects' samples obtained in MD simulations.